\documentclass[runningheads]{llncs}
\usepackage{graphicx}
\usepackage[utf8]{inputenc}
\usepackage{amsmath}
\usepackage{bbm}
\usepackage{xcolor}
\usepackage{hyperref}
\usepackage{amssymb}
\everymath{\displaystyle}
\begin{document}

\title{Conformal Coordinates for Molecular Geometry: from 3D to 5D}

%

\author{Jesus Camargo \inst{1}, Carlile Lavor\inst{2}, Michael Souza\inst{3}}

\institute{CCET, The Western Paraná State University, 85819-110, Cascavel, Brazil\\\email{jesus.m.camargo@outlook.com }\and
IMECC, University of Campinas, 13081-970, Campinas, Brazil\\\email{clavor@unicamp.br} \and
Departamento de Estatística e Matemática Aplicada, Centro de Ciências, Universidade Federal do Ceará, 60020-181, Fortaleza, Brazil,\\\email{michael@ufc.br}}

%
\authorrunning{J. Camargo et al.}
%
%
\maketitle              
\begin{abstract}

This paper introduces the conformal model (an extension of the homogeneous coordinate system) for molecular geometry, where 3D space is represented within $\mathbb{R}^5$ with an inner product different from the usual one. This model enables efficient computation of interatomic distances using what we call the \textit{Conformal Coordinate Matrix} (\textit{C-matrix}). The \textit{C-matrix} not only simplifies the mathematical framework but also reduces the number of operations required for distance calculations compared to traditional methods. 

\end{abstract}

\section{Introduction}


In computational chemistry, the geometric arrangement of atoms within a molecule is often represented using Cartesian or internal coordinates (given by the lengths of covalent bonds and the bond and torsion angles), which are particularly useful because they are closely related to the chemical bonds and angles that define the molecule's structure \cite{Li_23}.

The traditional approach to converting internal coordinates into Cartesian coordinates involves the use of the homogeneous coordinate system. In this system, each point in 3D space is represented by a vector in $\mathbbm{R}^4$, allowing for translation and rotation of atoms to be described by matrix operations. This method, as proposed by Thompson in the 1960s \cite{Thompson}, has been widely used in molecular geometry calculations (for example, see \cite{lavor_14}).

While the homogeneous coordinate system simplifies the conversion of internal coordinates to Cartesian coordinates, it does not inherently simplify the calculation of interatomic distances, which is a crucial task in molecular geometry optimization and molecular dynamics simulations \cite{parsons_05,rybkin_13}. To address this limitation, this paper introduces the conformal model for molecular geometry, a generalization of the homogeneous coordinate system \cite{dress_93,hestenes_01,li_01}.

In the conformal model, we define the \textit{Conformal Coordinate Matrix} (\textit{C-matrix}), which allows for a more efficient computation of interatomic distances. The \textit{C-matrix} not only retains the advantages of the homogeneous coordinate system but also introduces a new level of computational efficiency by reducing the number of operations required for distance calculations. 

This paper explores the mathematical framework of the conformal model, demonstrates its application to molecular geometry, and compares its performance with traditional methods. 

\section{Homogeneous coordinate system}

\begin{figure}
    \centering
    \includegraphics[width=0.4\linewidth]{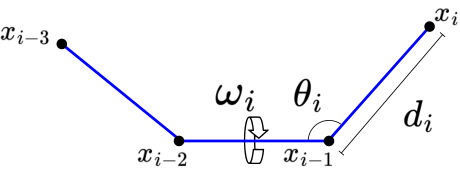}
    \caption{Cartesian and internal coordinates}
    \label{fig:enter-label}
\end{figure}

Since internal coordinates are naturally associated with the geometry of a molecule, especially when bond lengths and bond angles are considered fixed and given a priori (which reduces the degrees of freedom needed to characterize the 3D structure of a molecule), internal coordinates are widely used in computational chemistry \cite{parsons_05} (see Fig. \ref{fig:enter-label}).

We will consider, then (as in \cite{camargo_24}), a molecule as a linear chain of \( n \) atoms described by internal coordinates \( d_i, \theta_i, \omega_i \), where \( d_i \) is the covalent bond length between atoms with Cartesian coordinates $x_{i-1}, x_i \in \mathbbm{R}^3$ ($i=2,\ldots, n$), \( \theta_{i} \) is the angle formed by the bond vectors \( b_{i-1}, b_i \), given by \( b_i = x_{i} - x_{i-1} \) ($i=3,\ldots, n$), and \( \omega_i \) is the torsion angle formed by the planes generated by \( b_{i-2}, b_{i-1} \) and \( b_{i-1}, b_i \) ($i=4,\ldots, n$).

In \cite{Thompson} (see also \cite{Thompson_2}), Thompson proposes using the homogeneous model of 3D space (where each point is represented by $(x,y,z,1)^t \in \mathbbm{R}^4$) to convert internal coordinates into Cartesian coordinates. This approach allows for grouping the three ``positioning movements" of an atom (one translation and two rotations), considering previous atoms in the chain, into a single linear operator. As a result, the calculation of Cartesian coordinates from internal coordinates is simply given by a matrix product, as we summarize below following the procedure given in \cite{camargo_24}.

The rotations associated with the bond and torsion angles are given in the homogeneous space by
\begin{align*}
B_{\theta_i}=\begin{bmatrix}
-\cos \theta_i & -\sin \theta_i & 0 & 0\\
\sin \theta_i  & -\cos \theta_i & 0 & 0\\
0 & 0 & 1 & 0\\
0 & 0 & 0 & 1
\end{bmatrix} \quad\text{and}\quad
B_{\omega_i}=\begin{bmatrix}
1 & 0 & 0 & 0\\
0 & \cos\omega_i & -\sin\omega_i & 0\\
0 & \sin\omega_i & \cos\omega_i & 0\\
0 & 0 & 0 & 1
\end{bmatrix},
\end{align*}
respectively, and the translation of a point $x_i\in\mathbbm{R}^3$ is encoded by
\begin{align*} 
\left[\begin{array}{cccc}
1 & 0 & 0 & d_i\\
0 & 1 & 0 & 0\\
0 & 0 & 1 & 0\\
0 & 0 & 0 & 1\\
\end{array}\right].
\end{align*}

Combining these matrices, we get
\begin{small}
\begin{eqnarray*}
B_i &=& \begin{bmatrix}
1 & 0 & 0 & 0\\
0 & \cos\omega_i & -\sin\omega_i & 0\\
0 & \sin\omega_i & \cos\omega_i & 0\\
0 & 0 & 0 & 1
\end{bmatrix}\begin{bmatrix}
-\cos \theta_i & -\sin \theta_i & 0 & 0\\
\sin \theta_i  & -\cos \theta_i & 0 & 0\\
0 & 0 & 1 & 0\\
0 & 0 & 0 & 1
\end{bmatrix}\begin{bmatrix}
1 & 0 & 0 & d_i\\
0 & 1 & 0 & 0\\
0 & 0 & 1 & 0\\
0 & 0 & 0 & 1\\
\end{bmatrix}\\
&=& \begin{bmatrix}
-\cos \theta_i & -\sin \theta_i & 0 & -d_i\cos\theta_i\\
\sin \theta_i \cos\omega_i & -\cos \theta_i\cos\omega_i & -\sin\omega_i & d_i\sin\theta_i\cos\omega_i\\
\sin\theta_i\sin \omega_i & -\cos\theta_i\sin\omega_i & \cos\omega_i & d_i\sin\theta_i\sin\omega_i\\
0 & 0 & 0 & 1
\end{bmatrix}. 
\end{eqnarray*}
\end{small}
For \(d_1=\omega_1=\omega_2=\omega_3=0\) and \(\theta_1=\theta_2=\pi\), we obtain

\begin{small}
\begin{align}
x_1 &= B_1e_4 = \left[ \begin{array}{cccc}
0\\
0\\
0\\
1
\end{array} \right],\,\,
x_2 = (B_1B_2)e_4 = \left[ \begin{array}{cccc}
d_2\\
0\\
0\\
1
\end{array} \right],\,\,
x_3  = (B_1B_2B_3)e_4 = \left[ \begin{array}{cccc}
(d_2-d_3\cos\theta_3)\\
d_3\sin\theta_3\\
0\\
1
\end{array} \right],\nonumber
\end{align}
\end{small}
and, for $i=4,\ldots,n$, 
\begin{equation*}
  x_i  = (B_4 \cdots B_i)e_4,
\end{equation*}
where $e_4 = (0,0,0,1)^t$ is the vector in the homogeneous space that represents the ``origin'' of 3D space.

To simplify the notation, let us write $$B_{[i,j]}=\prod_{k=i}^jB_k$$ and calculate the Euclidean distance $r_{i,j}$ between $x_i$ and $x_j$ (for $i<j$):
\begin{eqnarray*}
r_{i,j}&=&\Vert(x_j-x_i)\Vert\\
&=&\Vert (B_1\cdots B_i\cdots B_j)e_4-(B_1\cdots B_i)e_4\Vert\\
&=& 
\left\Vert B_{[1,i]}\left(B_{[i+1,j]}-I\right)e_4\right\Vert,
\end{eqnarray*}
where $I$ is the identity matrix in $\mathbbm{R}^{4\times 4}$.

Although the term $B_{[1,i]}$ is not an orthogonal matrix, the authors in \cite{camargo_24} show that it can be removed, resulting in
\begin{eqnarray*}
r_{i,j}&=&\left\Vert\left(B_{[i+1,j]}-I\right)e_4\right\Vert.
\end{eqnarray*}
\section{Conformal coordinate system}
To obtain the position of the fourth atom in the molecule, $$x_4 = (B_1 B_2 B_3 B_4)e_4,$$ we first calculate $B_4 e_4$, given by 
\begin{small}
\begin{eqnarray*}
B_4e_4=
\begin{bmatrix}
-\cos \theta_4 & -\sin \theta_4 & 0 & -d_4\cos\theta_4\\
\sin \theta_4 \cos\omega_4 & -\cos \theta_4\cos\omega_4 & -\sin\omega_4 & d_4\sin\theta_4\cos\omega_4\\
\sin\theta_4\sin \omega_4 & -\cos\theta_4\sin\omega_4 & \cos\omega_4 & d_4\sin\theta_4\sin\omega_4\\
0 & 0 & 0 & 1
\end{bmatrix}\begin{bmatrix}
0 \\
0 \\
0 \\
1\\
\end{bmatrix}=\begin{bmatrix}
-d_4\cos\theta_4 \\
d_4\sin\theta_4\cos\omega_4 \\
d_4\sin\theta_4\sin\omega_4 \\
1\\
\end{bmatrix}.
\end{eqnarray*}
\end{small} 

In the above calculation, we see that the rotations associated with the bond angle $\theta_4$ and the torsion angle $\omega_4$ are represented by
\begin{align*}
A=\begin{bmatrix}
-\cos \theta_4 & -\sin \theta_4 & 0\\
\sin \theta_4 \cos\omega_4 & -\cos \theta_4\cos\omega_4 & -\sin\omega_4\\
\sin\theta_4\sin \omega_4 & -\cos\theta_4\sin\omega_4 & \cos\omega_4\\
\end{bmatrix}
\end{align*}
and the translation associated with the bond length $d_4$ (applied to a unit vector that has already undergone two rotations by the angles \( \theta_4 \) and \( \omega_4 \)) is encoded by
\begin{small}
\begin{eqnarray*}
b &=& \begin{bmatrix}
-d_4\cos\theta_4\\
d_4\sin\theta_4\cos\omega_4\\
d_4\sin\theta_4\sin\omega_4\\
\end{bmatrix}.
\end{eqnarray*}
\end{small}

In other words, using homogeneous coordinates, the translation (in 3D space) is linearized (in 4D space) and the three operations are represented by a single matrix. This linearization, in general, can be represented by
\begin{align}
\begin{bmatrix}
A & b\\
0 & 1 \\
\end{bmatrix} \begin{bmatrix}
x\\
1\\
\end{bmatrix}=\begin{bmatrix}
Ax + b \\
1 \\
\end{bmatrix},
\label{Ab}
\end{align}
where $x \in \mathbbm{R}^3$.

Note that the matrix $A$, related to the rotations, is orthogonal. However, when we linearize the translation, the new matrix, now in $\mathbbm{R}^{4\times 4}$, is no longer orthogonal.

In \cite{lavor_21}, the authors manage to ``recover" this property (slightly modifying the concept of orthogonality) by using another model of 3D space, called the \textit{conformal model} \cite{dress_93,hestenes_01,li_01}.

In $\mathbbm{R}^{3}$, the two rotations and the translation given in (\ref{Ab}) can be represented by a function $f: \mathbbm{R}^3 \to \mathbbm{R}^3$, defined by
$$f(x)=Ax+b,$$
where $A \in\mathbbm{R}^{3\times 3}$, such that $A^{-1}=A^t$, and $b \in\mathbbm{R}^{3}$. That is, $f$ is an isometry in $\mathbbm{R}^{3}$.

Also in \cite{lavor_21}, it is demonstrated that it is not possible to ``orthogonalize" isometries in 3D space using the homogeneous model. However, by renouncing the positivity of the usual inner product and using the conformal model, one can encode translations in 3D space as orthogonal operations in $\mathbbm{R}^{5}$. In that paper, the motivation was the search for an orthogonal representation of isometries in 3D space. Perhaps, due to the chosen notation, the development of this reasoning was not very clear. We present an alternative below, which we believe is more convincing.

\subsection{Orthogonalization of isometries}

The entire argument in \cite{lavor_21} is based on constructing a bijection between $\mathbbm{R}^{3}$ and a subset $\mathbbm{H} \subset \mathbbm{R}^{5}$, in such a way that isometries in $\mathbbm{R}^{3}$ can be represented orthogonally in $\mathbbm{R}^{5}$. For each $x \in \mathbbm{R}^{3}$, its representative in $\mathbbm{H}$ will be denoted by $\hat{x} \in \mathbbm{R}^{5}$.

We want that when applying an isometry $f: \mathbbm{R}^3 \to \mathbbm{R}^3$ to $x,y\in \mathbbm{R}^{3}$, their respective representatives $\hat{x},\hat{y} \in \mathbbm{H} \subset \mathbbm{R}^5$ are altered orthogonally.

In other words, we would like to demonstrate that, for any $x,y \in\mathbbm{R}^3$,  
$$\Vert x-y \Vert=\Vert f(x)-f(y)\Vert \implies \hat{x} \cdot \hat{y}=\widehat{f(x)} \cdot \widehat{f(y)}.$$

 One way to obtain this result would be to assume that the inner product in $\mathbbm{R}^{5}$ ``encodes'' the Euclidean distance in $\mathbbm{R}^{3}$. That is, if
$$\hat{x} \cdot \hat{y}=\Vert x-y\Vert ^2,$$
we would have 
$$ \widehat{f(x)} \cdot \widehat{f(y)}=\Vert f(x)-f(y)\Vert ^2,$$
which would imply 
$$\hat{x} \cdot \hat{y}=\widehat{f(x)} \cdot \widehat{f(y)}.$$

The question, therefore, is to investigate how the hypothesis $\hat{x} \cdot \hat{y}=\Vert x-y\Vert ^2$, with $x,y \in \mathbbm{R}^{3}$ and $\hat{x},\hat{y} \in \mathbbm{R}^5$, could lead to the discovery of the bijection in question between $\mathbbm{R}^{3}$ and the subset $\mathbbm{H} \subset \mathbbm{R}^5$.

So far, we have considered the usual inner product, both in $\mathbbm{R}^3$ and in $\mathbbm{R}^5$, which induces Euclidean norms in both spaces.

The first consequence of the hypothesis $\hat{x} \cdot \hat{y}=\Vert x-y\Vert ^2$ is that, setting $x=y$, we would have
$$(\forall x \in\mathbbm{R}^3, \quad \hat{x} \cdot \hat{x}=0) \implies   \| \hat{x} \| =0\implies \hat{x}=0,$$
which would be a contradiction because we are looking for a bijection.

To avoid this contradiction, we will abandon the positivity of the inner product in $\mathbbm{R}^5$, which implies that (assuming $\hat{x} \cdot \hat{y}=\Vert x-y\Vert ^2$), 
$$(\forall x \in\mathbbm{R}^3, \quad \hat{x} \cdot \hat{x}=0) \quad \implies \quad \| \hat{x} \| =0.$$

That is, we will admit that all points in 3D space will be represented by vectors in $\mathbbm{R}^{5}$ with zero norm. Of course, this norm, induced by the new inner product, will no longer be Euclidean (we will continue to call it ``the new inner product", even though we know that, formally, we no longer have such operation due to the lack of positivity).

The other properties that define an inner product will be preserved. This means that we will maintain the algebraic properties of the inner product in $\mathbbm{R}^{5}$ (symmetry, homogeneity, and distributivity), but as the Euclidean character requires the positivity of the inner product, the geometry in $\mathbbm{R}^{5}$ will be altered. Therefore, we are looking for a non-Euclidean representation (that lies in $\mathbbm{R}^{5}$) for the 3D space. By abuse of notation, we will continue writing $\hat{x}\cdot\hat{y}$ to represent the new inner product.

From \cite{lavor_21}, knowing that it is not possible to orthogonalize isometries of 3D space in $\mathbbm{R}^4$, even by relinquishing the positivity of the inner product, we can follow what is done in the homogeneous model and represent a point $x=x_1e_1+x_2e_2+x_3e_3 \in \mathbbm{R}^3$ in $\mathbbm{H} \subset \mathbbm{R}^5$ by 
$$\hat{x}=x+x_4e_4 + x_5e_5,$$
where $x_1,x_2,x_3,x_4,x_5 \in \mathbbm{R}$ and, together with ${e_1,e_2,e_3} \in \mathbbm{R}^{3}$, ${e_4,e_5}$ are vectors that complete the canonical basis of $\mathbbm{R}^5$. Of course, for the above sum to make sense, we add zeros to the fourth and fifth coordinates when embedding $x,e_1,e_2,e_3$ in $\mathbbm{R}^5$. The problem now is to determine the values of $x_4$ and $x_5$. 

From the algebraic properties of the inner product in $\mathbbm{R}^5$, we easily obtain that 
\begin{small}
$$\hat{x} \cdot \hat{x}=0 \implies (x+x_4e_4+x_5e_5) \cdot (x+x_4e_4+x_5e_5)=0\implies x_4^2(e_4 \cdot e_4)+ x_5^2(e_5 \cdot e_5)=- \Vert x \Vert^2.$$
\end{small}
In other words,
$$(x \neq 0 \quad \text{and} \quad \Vert e_4 \Vert =1) \implies \Vert e_5 \Vert < 0.$$ 
Because the norm in $\mathbbm{R}^5$ is no longer Euclidean, negative norm, as well as zero norm of a non-zero vector, is no longer forbidden. 

Let us also note that the set $\{e_1,e_2,e_3,e_4,e_5\}$ is orthogonal, but no longer orthonormal because we will assume $\Vert e_5 \Vert = -1$.

Since the vectors in $\mathbbm{H} \subset \mathbbm{R}^5$, which represents the points in 3D space, must have zero norm, we will replace $e_4,e_5$ with vectors $e_0,e_\infty$ that also have zero norm (see \cite{lavor_21} for more details), defined by
$$e_0= \frac{e_5-e_4}{2} \quad \text{and} \quad e_{\infty}=e_5 + e_4,$$
resulting in
$$\Vert e_0 \Vert = \Vert e_{\infty} \Vert =0$$
and
$$e_0 \cdot e_{\infty}=-1.$$

With the new basis $\{e_1,e_2,e_3,e_0,e_\infty\}$, and given that $e_0,e_\infty$ are also orthogonal to $\{e_1,e_2,e_3\}$, we obtain
\begin{equation}
    \hat{x} \cdot \hat{y}=(x + x_0e_0 + x_{\infty}e_{\infty}) \cdot (y + y_0e_0 + y_{\infty}e_{\infty})=x \cdot y - (x_0y_{\infty} + x_{\infty}y_0),
    \label{xy} 
\end{equation}
with $x,y \in \mathbbm{R}^3$, $x_0,x_\infty \in \mathbbm{R}$, and $e_0,e_\infty \in \mathbbm{R}^5$.

For $\hat{x} = \hat{y}$,
$$\hat{x} \cdot \hat{x}=0 \implies \Vert x \Vert ^2 - 2x_0x_{\infty}=0$$
and, considering $x_0=1$, we finally obtain
\begin{equation}
    \hat{x} = x+e_0+ \frac{1}{2} \Vert x \Vert ^2 e_{\infty},\label{m_c} 
\end{equation}
implying that 
$$\hat{x} \cdot \hat{y}= \left ( x+e_0+ \frac{1}{2} \Vert x \Vert ^2 e_{\infty} \right) \cdot \left ( y+e_0+ \frac{1}{2} \Vert y \Vert ^2 e_{\infty} \right) = - \frac{1}{2} \Vert x-y \Vert ^2.$$

The expression \eqref{m_c} defines the conformal model \cite{hestenes_01,li_01}. Since $e_0$ represents the ``origin" of 3D space $(\hat{x} = x+e_0+ 0.5 \Vert x \Vert ^2 \implies \hat{0} = e_0)$, the conformal model is also known as the generalized homogeneous model (see \cite{lavor_21} for details). 



With the bijection between $\mathbbm{R}^{3}$ and the subset $\mathbbm{H} \subset \mathbbm{R}^5$, defined by \eqref{m_c}, our hypothesis becomes true with a slight adjustment. That is, for $x,y \in \mathbbm{R}^3$,
$$\hat{x} = x+e_0+ \frac{1}{2} \Vert x \Vert ^2 e_{\infty} \implies \hat{x} \cdot \hat{y}=  - \frac{1}{2} \Vert x-y \Vert ^2,$$
which, in turn, implies that

$$\Vert x-y \Vert=\Vert f(x)-f(y)\Vert \implies \hat{x} \cdot \hat{y}=\widehat{f(x)} \cdot \widehat{f(y)}.$$

Thus, we reach our goal: when applying an isometry $f: \mathbbm{R}^3 \to \mathbbm{R}^3$ to $x,y \in \mathbbm{R}^{3}$, their respective representatives $\hat{x},\hat{y} \in \mathbbm{R}^5$ are altered orthogonally (of course, considering the new inner product in $\mathbbm{R}^5$, which no longer respects positivity).

\subsection{Conformal coordinate matrix (\textit{C-matrix})}

Using the conformal model, an isometry $f: \mathbbm{R}^3 \to \mathbbm{R}^3$,
$$f(x)=Ax+b,$$
is then represented by
$$\widehat{f(x)}=(Ax+b)+e_0+\frac{1}{2} \Vert Ax+b \Vert ^2 e_{\infty}.$$
Because of the orthogonality of the matrix $A$, we get $$\frac{1}{2} \Vert Ax+b \Vert ^2 = b^tAx + \frac{\Vert b \Vert ^2}{2} + \frac{\Vert x \Vert ^2}{2},$$
which implies that the isometry $f$ can be encoded in matrix form as
\begin{align*}
\begin{bmatrix}
A & b & 0\\
0 & 1 & 0\\
b^tA & \frac{\Vert b \Vert ^2}{2} & 1\\
\end{bmatrix} \begin{bmatrix}
x\\
1\\
\frac{\Vert x \Vert ^2}{2}\\
\end{bmatrix}=\begin{bmatrix}
Ax + b \\
1 \\
\frac{\Vert Ax+b \Vert ^2}{2} \\
\end{bmatrix},
\end{align*}
where $x \in \mathbbm{R}^3$ (see details in \cite{lavor_21}).

Taking
\begin{align*}
A=\begin{bmatrix}
-\cos \theta_i & -\sin \theta_i & 0\\
\sin \theta_i \cos\omega_i & -\cos \theta_i\cos\omega_i & -\sin\omega_i\\
\sin\theta_i\sin \omega_i & -\cos\theta_i\sin\omega_i & \cos\omega_i\\
\end{bmatrix}
 \quad \text{and}\quad 
b = \begin{bmatrix}
-d_i\cos\theta_i\\
d_i\sin\theta_i\cos\omega_i\\
d_i\sin\theta_i\sin\omega_i\\
\end{bmatrix},
\end{align*}
we have
\begin{align*}
b^tA&=\begin{bmatrix}
    d_i && 0 && 0
\end{bmatrix} \quad\text{and}\quad\Vert b \Vert ^2 = d_i^2,
\end{align*}
implying that
\begin{align*}
\begin{bmatrix}
A & b & 0\\
0 & 1 & 0\\
b^tA & \frac{\Vert b \Vert ^2}{2} & 1\\
\end{bmatrix} = 
\begin{bmatrix}
-\cos \theta_i & -\sin \theta_i & 0 & -d_i\cos\theta_i & 0\\
\sin \theta_i \cos\omega_i & -\cos \theta_i\cos\omega_i & -\sin\omega_i & d_i\sin\theta_i\cos\omega_i & 0\\
\sin\theta_i\sin \omega_i & -\cos\theta_i\sin\omega_i & \cos\omega_i & d_i\sin\theta_i\sin\omega_i & 0\\
0 & 0 & 0 & 1 & 0\\
d_i & 0 & 0 & \frac{d_i^2}{2} & 1
\end{bmatrix}.
\end{align*}

We denote this matrix as the \textit{Conformal Coordinate Matrix} of atom $i$ or simply the \textit{C-matrix} of atom $i$.

The \textit{C-matrix} is not orthogonal with respect to the usual inner product, but it is if we consider the new one. That is, it satisfies
$$(U\hat{x})\cdot(U\hat{y})=\hat{x}\cdot\hat{y},$$ as we can see in what follows.

From (\ref{xy}),
$$\hat{x} \cdot \hat{y}= x_1y_1 + x_2y_2 + x_3y_3 - x_0y_{\infty} - x_{\infty}y_0,$$
which implies that, in matrix format, 
\begin{equation}
\hat{x} \cdot \hat{y}=\begin{bmatrix}
    x_1 && x_2 && x_3 && x_0 && x_{\infty} 
\end{bmatrix} 
\begin{bmatrix}
1 & 0 & 0 & 0 & 0\\
0 & 1 & 0 & 0 & 0\\
0 & 0 & 1 & 0 & 0\\
0 & 0 & 0 & 0 & -1\\
0 & 0 & 0 & -1 & 0\\
\end{bmatrix} 
\begin{bmatrix}
y_1 \\
y_2 \\
y_3 \\
y_0 \\
y_{\infty}\\
\end{bmatrix}
=\hat{x}^tI_c \hat{y},
\label{x_y}
\end{equation}
where
$$I_c = 
\begin{bmatrix}
I & 0 & 0\\
0 & 0 & -1\\
0 & -1 & 0 
\end{bmatrix}$$
and $I \in \mathbbm{R}^{3\times 3}$. Considering
\begin{align*}
U= 
\begin{bmatrix}
A & b & 0\\
0 & 1 & 0\\
b^tA & \frac{\Vert b \Vert ^2}{2} & 1\\
\end{bmatrix},
\end{align*}
we obtain
\begin{equation}
    U^tI_cU=I_c
    \label{I_c}
\end{equation}
and, in turn,

$$(U\hat{x})\cdot(U\hat{y})=(U\hat{x})^tI_c(U\hat{y})=\hat{x}^t(U^tI_cU)\hat{y}=\hat{x}\cdot\hat{y},$$
for all $x,y \in \mathbbm{R}^3$.

Thus, $U$ (and, in particular, the \textit{C-matrix}) is an orthogonal matrix with respect to the inner product defined by \eqref{x_y}.

Finally, we are ready to compute distances using the conformal model.

\section{Computing distances in the conformal coordinate system}

Since the computation of Cartesian coordinates from internal coordinates (using the conformal model) follows the same procedure performed in the homogeneous space (see Section 2), we have that 
$$\hat{x}_i=B_{[i]}e_0,$$
where  
\begin{align*}
B_{[i]}=\prod_{k=1}^iB_k,\quad\quad \hat{x}_i = 
\begin{bmatrix}
x_i\\
1\\
\frac{\Vert x_i \Vert ^2}{2}\\
\end{bmatrix},
\end{align*}
and
\begin{align*}
B_i= 
\begin{bmatrix}
-\cos \theta_i & -\sin \theta_i & 0 & -d_i\cos\theta_i & 0\\
\sin \theta_i \cos\omega_i & -\cos \theta_i\cos\omega_i & -\sin\omega_i & d_i\sin\theta_i\cos\omega_i & 0\\
\sin\theta_i\sin \omega_i & -\cos\theta_i\sin\omega_i & \cos\omega_i & d_i\sin\theta_i\sin\omega_i & 0\\
0 & 0 & 0 & 1 & 0\\
d_i & 0 & 0 & \frac{d_i^2}{2} & 1
\end{bmatrix},
\end{align*}
with $x_i \in \mathbbm{R}^3$.

As mentioned earlier, $e_0$ is the representative of the origin of 3D space in the conformal model, playing the role of $e_4$ in the homogeneous model.

Without loss of generality, let us consider $i<j$. Also defining
$$B_{[i+1,j]}=\prod_{k=i+1}^jB_k$$
and writing
$$B_{[j]}=B_{[i]}B_{[i+1,j]},$$
we obtain
\begin{align*}
 \hat{x}_j\cdot \hat{x}_i&=\hat{x}_j^tI_c \hat{x}_i\\
 &=e_0^tB_{[j]}^tI_cB_{[i]}e_0\\
&=e_0^tB_{[i+1,j]}^t(B_{[i]}^tI_cB_{[i]})e_0
\end{align*} 
and, from \eqref{I_c},
\begin{align*}
 \hat{x}_j\cdot \hat{x}_i&=e_0^tB_{[i+1,j]}^tI_ce_0\\
&=-e_0^tB_{[i+1,j]}^te_\infty.
\end{align*}
Since $\hat{x}_j\cdot \hat{x}_i \in \mathbbm{R}$,
$$(-e_0^tB_{[i+1,j]}^te_\infty)^t=-e_\infty^tB_{[i+1,j]}e_0,$$
which implies
$$\hat{x}_j\cdot \hat{x}_i=-e_\infty^tB_{[i+1,j]}e_0.$$
As we know that 
$$\hat{x}_j\cdot \hat{x}_i=  - \frac{1}{2} \Vert x_j - x_i \Vert ^2,$$
we have that the Euclidean distance $r_{i,j}$ between atoms $i$ and $j$ is given by
\begin{equation*}
    r_{i,j}^2=2e_\infty^tB_{[i+1,j]}e_0. 
\end{equation*}
Comparing this with the expression obtained using the homogeneous model \cite{camargo_24}, given by
\begin{align*}
r_{i,j}^2 &= e^t_4(B_{[i+1,j]}^tB_{[i+1,j]})e_4-1,
\end{align*}
we can see that the simplification obtained is due to the orthogonality of $B_{[i]}$, a consequence of the orthogonality of the \textit{C-matrix} (of course, orthogonality in terms of the new inner product in $\mathbbm{R}^5$). We can conclude this because
$$(B_iB_j)^tI_c(B_iB_j)=B_j^t(B_i^tI_cB_i)B_j=B_j^tI_cB_j=I_c.$$

\subsection{Number of operations for computing $r_{i,j}$}

In \cite{camargo_24}, a comparison was made between the Euclidean and homogeneous models regarding the number of operations (additions and multiplications) required to calculate the interatomic distance $r_{i,j}$ between atoms $i$ and $j$. As in \cite{camargo_24}, we will disregard the cost associated with calculating sine and cosine functions, as well as the square root, since they appear in equal numbers in all three models.

To compute $r_{i,j}$ in the conformal model, we need to calculate 
$$r_{i,j}=\sqrt{2e_{\infty}^tB_{[i+1,j]}e_0}.$$

First, let us determine separately the cost associated with the vectors $e_\infty^tB_{i+1}$ and $B_{[i+2,j]}e_0$.

Note that the first of these two vectors is exactly given by the fifth row of $B_{i+1}$, which requires only 2 multiplications.

For the second vector, $B_{[i+2,j]}e_0$, we need to calculate the fourth column of the matrix resulting from the product $B_{i+2}B_{i+3} \cdots B_{j}$. We will do this through a sequence of matrix-vector multiplications, operating from right to left.

The first vector calculated is $B_je_0$, which is given by the fourth column of $B_j$ and requires 5 multiplications, as $(d_i^2/2)$ has already been calculated in the first vector, $e_\infty^tB_{i+1}$. The sequence of matrix-vector multiplications is performed such that, for $p=i+2, \ldots,j-1$, we need to calculate
$$B_p\left(B_{[p+1,j]}e_0\right).$$

The cost of determining $B_p$ is 9 multiplications and the cost of multiplying the matrix $B_p$ by the vector $B_{[p+1,j]}e_0$ is 25 multiplications and 20 additions, totaling 54 operations for each index $p$. 

Disregarding the count of multiplications by 0 and 1, and additions with 0, the product of a matrix $B_p$ and a vector, whose fourth component is always equal to 1, requires 9 multiplications and 10 additions. Along with the 9 operations needed to determine each matrix $B_p$, computing the vector $B_{[p+1,j]}e_0$ requires 28 operations, for $p=i+2, \ldots,j-1$. Considering the 5 multiplications to determine $B_je_0$ and the 2 multiplications to determine $e_\infty^tB_{[i+1]}$, we have, so far, $28(j-i-2)+7$ operations.

The product between $e_\infty^tB_{i+1}$ and $B_{[i+2,j]}e_0$ requires 1 multiplication and 2 additions (considering that $e_\infty^t B_{i+1}$ has two zero entries and one entry equal to one, and that $B_{[i+2,j]}e_0$ has one entry equal to one) and we also have to consider the product of the resulting value by two. Thus, the total number of operations to calculate $r_{i,j}$, using the conformal model, is 
$$28(j-i-2)+7+4=28(j-i)-45.$$

In Table \ref{tab}, we compare the cost determined here with the cost obtained using the homogeneous and Euclidean models, as given in \cite{camargo_24} (note that, to make sense, we assume that $j>i+2$).
\begin{table}[h!]
\centering
\caption{Number of operations required to determine $r_{i,j}$, for $j>i+2$, with the Euclidean, homogeneous, and conformal models.}
  \begin{tabular}[htbp]{@{}lc@{}}
    \hline
    Model & Number of Operations\\
    \hline
    Euclidean  & $55(j-i)-97$ \\
    Homogeneous  & $35(j-i)-25$   \\
    Conformal & $28(j-i)-45$\\
    \hline
  \end{tabular}
  \label{tab}
\end{table}

\section{Conclusion}
The conformal model is a generalization of the homogeneous coordinate system, where 3D space is represented by a subset of $\mathbbm{R}^{5}$ with an inner product that no longer respects positivity. Widely used in problems of robotics, physics, and computer graphics \cite{bayro-corrochano_19,doran_03,dorst_07,kanatani_15}, this paper applies the conformal model, for the first time (as far as we know), to represent atomic positions and calculate interatomic distances in the context of molecular geometry.

As a result, we define the \textit{C-matrix}, which, like the \textit{Z-matrix}\footnote{https://en.wikipedia.org/wiki/Z-matrix\_(chemistry)}, encodes the geometry of a molecule in terms of internal coordinates. However, while the Z-matrix is primarily used as a structured representation of molecular configurations, the C-matrix leverages the conformal model to enable a computationally more efficient calculation of interatomic distances.

\subsection*{Data Availability Statement} \par 

Data sharing is not applicable to this article as no new data were created or analyzed in this study.

\subsection*{Conflict of Interest Statement} \par 

The authors declare no conflicts of interest.

\subsection*{Acknowledgements} \par 

This research was partially funded by the Brazilian research agencies FAPESP (grant numbers 2013/07375-0, 2023/08706-1, 2024/00923-6) and CNPq (grant numbers 305227/2022-0, 	404616/2024-0, 402609/2025-5). We would also like to thank Prof. Alberto Saa and Prof. Marcelo Terra Cunha, both from the University of Campinas, for the fruitful discussions on all the ideas in the paper.

%
%
%
 \bibliographystyle{splncs04}
 \bibliography{mybibliography}

@article{lavor_21,
  title={Orthogonality of isometries in the conformal model of the 3D space},
  author={Lavor, Carlile and Souza, Michael and Arag{\'o}n, Jos{\'e} Luis},
  journal={Graphical Models},
  volume={114},
  pages={101100},
  year={2021},
  publisher={Elsevier}
}

@article{Parsons_05,
  title={Practical conversion from torsion space to Cartesian space for in silico protein synthesis},
  author={Parsons, J. and Holmes, J. and Rojas, J. and Tsai, J. and Strauss, C.},
  journal={Journal of Computational Chemistry},
  volume={26},
  pages={1063--1068},
  year={2005}
}

@article{camargo_24,
  title={A new perspective on the homogeneous coordinate system for calculating interatomic distances and their derivatives in terms of internal coordinates},
  author={Camargo, Jesus and Lavor, Carlile},
  journal={Advanced Theory and Simulations},
  volume={7},
  pages={2400271},
  year={2024}
}

@article{Thompson,
  title={Calculation of cartesian coordinates and their derivatives from internal molecular coordinates},
  author={Thompson, H Bradford},
  journal={The Journal of Chemical Physics},
  volume={47},
  pages={3407--3410},
  year={1967},
  publisher={AIP}
}

@article{Thompson_2,
  title={Calculation of cartesian coordinates and their derivatives from internal molecular coordinates. II. Second and higher derivatives of vectors},
  author={Thompson, H. Bradford},
  journal={The Journal of Chemical Physics},
  volume={53},
  pages={3034--3036},
  year={1970},
  publisher={AIP}
}

@book{bayro-corrochano_19,
  title={Geometric Algebra Applications. Computer Vision, Graphics and Neurocomputing I},
  author={Bayro-Corrochano, E.},
  publisher={Springer},
  year={2019}
}

@book{doran_03,
  title={Geometric Algebra for Physicists},
  author={Doran, C. and Lasenby, A.},
  publisher={Cambridge University Press},
  year={2003}
}

@book{dorst_07,
  title={Geometric Algebra for Computer Science: An Object-Oriented Approach to Geometry},
  author={Dorst, L. and Fontijne, D. and Mann, S.},
  publisher={Morgan Kaufman},
  year={2007}
}

@article{dress_93,
  title={Distance geometry and geometric algebra},
  author={Dress, A. and Havel, T.},
  journal={Found. Phys.},
  volume={23},
  pages={1357--1374},
  year={1993}
}

@incollection{hestenes_01,
  title={Old wine in new bottles: a new algebraic framework for computational geometry},
  author={Hestenes, D.},
  booktitle={Advances in Geometric Algebra with Applications in Science and Engineering},
  editor={Bayro-Corrochano, E. and Sobczyk, G.},
  publisher={Birkhäuser},
  year={2001},
  pages={1--14}
}

@book{kanatani_15,
  title={Understanding Geometric Algebra: Hamilton, Grassmann, and Clifford for Computer Vision and Graphics},
  author={Kanatani, K.},
  publisher={CRC Press},
  year={2015}
}

@article{lavor_14,
  title={Euclidean distance geometry and applications},
  author={Liberti, Leo and Lavor, Carlile and Maculan, Nelson and Mucherino, Antonio},
  journal={SIAM Review},
  volume={56},
  pages={3-69},
  year={2014},
}

@article{Li_23,
  title={Learning correlations between internal coordinates to improve 3D Cartesian coordinates for proteins},
  author={Li, J. and Zhang, O. and Lee, S. and Namini, A. and Liu, Z.H. and Teixeira, J. and Forman-Kay, J. and Head-Gordon, T.},
  journal={Journal of Chemical Theory and Computation},
  volume={19},
  pages={4689-4700},
  year={2023},
}

@incollection{li_01,
  title={Generalized homogeneous coordinates for computational geometry},
  author={Li, H. and Hestenes, D. and Rockwood, A.},
  booktitle={Geometric Computing with Clifford Algebra},
  editor={Sommer, G.},
  publisher={Springer},
  year={2001},
  pages={25--58}
}

@article{rybkin_13,
  title={Internal-to-Cartesian back transformation of molecular geometry steps using high-order geometric derivatives},
  author={Rybkin, Vladimir and Ekström, Ulf and Helgaker, Trygve},
  journal={Journal of Computational Chemistry},
  volume={34},
  pages={1842-1849},
  year={2013},
}
\end{document}